\documentclass[sigconf]{acmart}

\usepackage{multirow}
\usepackage{siunitx}
\usepackage{enumitem}
\usepackage{tikz}
\usepackage{url}
\usepackage{ulem}

\newcommand{\commentout}[1]{}
\setlist[itemize]{leftmargin=\parindent,labelindent=0pt}
\setlist[description]{leftmargin=\parindent,labelindent=0pt}

\copyrightyear{2026}
\acmYear{2026}
\setcopyright{cc}
\setcctype{by}
\acmConference[UIST Adjunct '26]{The 39th Annual ACM Symposium on User Interface Software and Technology}{November 02--05, 2026}{Detroit, MI, USA}
\acmBooktitle{The 39th Annual ACM Symposium on User Interface Software and Technology (UIST Adjunct '26), November 02--05, 2026, Detroit, MI, USA}
\acmDOI{10.1145/3830397.3841874}
\acmISBN{979-8-4007-2855-6/2026/11}

\begin{document}

\title[Can People Distinguish Human and AI Agency in Humanoid Teleoperation?]{Can People Distinguish Human and AI Agency in Humanoid Teleoperation? A Preliminary Study of Agency Perception}

\author{Xiang Li}
\orcid{0000-0001-5529-071X}
\authornote{This work was conducted while Xiang Li was a research intern at Sony CSL – Kyoto and a guest researcher at The University of Tokyo.}
\authornote{These authors contributed equally to this work.}
\affiliation{%
  \institution{Sony Computer Science Laboratories}
  \city{Kyoto}
  \country{Japan}}
\affiliation{%
  \institution{University of Cambridge}
  \city{Cambridge}
  \country{United Kingdom}}
\email{xl529@cam.ac.uk}

\author{Koya Dendo}
\authornotemark[2]
\affiliation{%
  \institution{The University of Tokyo}
  \city{Tokyo}
  \country{Japan}}
\affiliation{%
  \institution{Sony Computer Science Laboratories}
  \city{Kyoto}
  \country{Japan}}
\email{koya-6732@g.ecc.u-tokyo.ac.jp}

\author{Keigo Minamida}
\authornotemark[2]
\affiliation{%
  \institution{The University of Tokyo}
  \city{Tokyo}
  \country{Japan}}
\affiliation{%
  \institution{Sony Computer Science Laboratories}
  \city{Kyoto}
  \country{Japan}}
\email{keigo-minamida@g.ecc.u-tokyo.ac.jp}

\author{Yuto Nakamura}
\affiliation{%
  \institution{The University of Tokyo}
  \city{Tokyo}
  \country{Japan}}
\email{yuto-nakamura@g.ecc.u-tokyo.ac.jp}

\author{Per Ola Kristensson}
\orcid{0000-0002-7139-871X}
\affiliation{%
\institution{University of Cambridge}
\city{Cambridge} 
\country{United Kingdom} 
}
\email{pok21@cam.ac.uk}

\author{Jun Rekimoto}
\affiliation{%
  \institution{Sony Computer Science Laboratories}
  \city{Kyoto}
  \country{Japan}}
  \affiliation{%
  \institution{The University of Tokyo}
  \city{Tokyo}
  \country{Japan}}
\email{rekimoto@acm.org}

\renewcommand{\shortauthors}{Xiang Li, Koya Dendo, Keigo Minamida, Yuto Nakamura, Per Ola Kristensson, and Jun Rekimoto}

\begin{abstract}

Can people distinguish between human and AI agency in humanoid teleoperation? To explore this question, we developed \textit{Ghost-in-the-Loop}, a teleoperation framework that supports both human-operated and AI-generated control of a robot's voice, facial expressions, and gestures while maintaining a consistent embodiment. We conducted a preliminary online study ($N=50$) in which participants viewed short interaction clips generated by either a Human Operator or an AI Control and judged the perceived source of control. Results suggest that participants often struggled to distinguish between the two conditions in brief interactions. Qualitative responses indicate that judgments were primarily influenced by perceived naturalness, temporal coordination, and consistency across speech, facial expression, and gesture. These findings provide initial insights into agency perception in embodied human--AI communication and motivate future investigations of blended human--AI telepresence systems.

\end{abstract}




\maketitle

\section{Introduction}

Humanoid teleoperation enables people to communicate through robotic embodiments that convey speech, facial expressions, and body movements, supporting applications such as remote collaboration, customer service, and caregiving~\cite{lee2011now,rae2012one}. However, conventional teleoperation requires continuous human involvement, limiting scalability and making it difficult for a single operator to support multiple users, robots, or interactions simultaneously.

Recent advances in generative AI have enabled increasingly realistic speech synthesis, voice cloning, facial animation, and gesture generation~\cite{nirkin2019fsgan,lavan2025voice,rebol2021real}. These developments raise the possibility that portions of teleoperated interaction could be delegated to AI systems while preserving the appearance of a consistent communication partner~\cite{floyd2026grand}. Rather than treating teleoperation and autonomy as separate operating modes, future embodied systems may dynamically transition between human- and AI-generated behavior according to context, workload, or availability.

At the same time, recent work has shown that AI-generated behavior is becoming increasingly difficult to distinguish from human behavior in individual modalities such as speech and embodied interaction~\cite{lavan2025voice,zhang2025react}. However, little is known about how users perceive agency when both human and AI behaviors are presented through the same embodied interface. In particular, it remains unclear whether observers can reliably distinguish between human-operated and AI-generated behavior in robot-mediated interaction, and which behavioral cues contribute to such judgments.

\begin{figure*}
    \centering
    \includegraphics[width=0.7\linewidth]{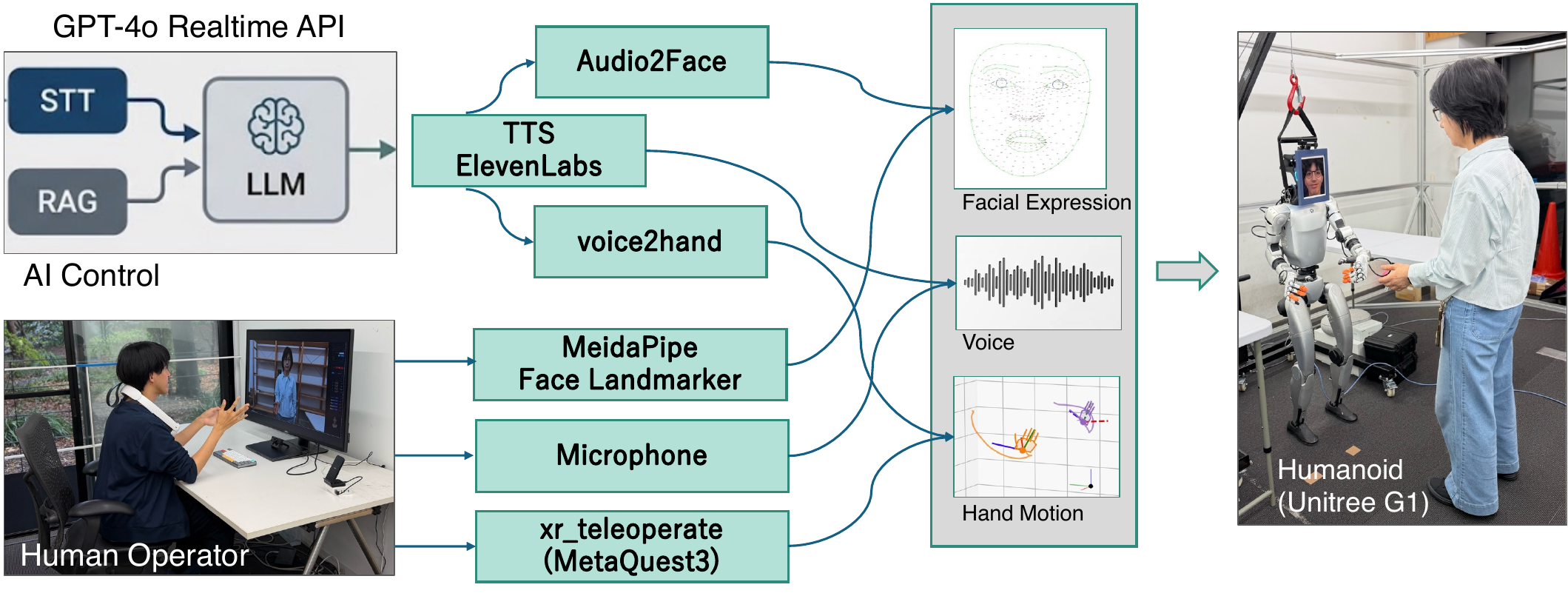}
    \caption{\textit{Ghost-in-the-Loop} architecture. Human teleoperation and AI-generated behaviors are presented through the same humanoid embodiment, sharing voice, facial expression, and gesture output channels.}
    \Description{A system architecture diagram showing AI and human control pipelines connected to a humanoid robot.}
    \label{fig:ghost_arch}
\end{figure*}

To explore this question, we developed \textit{Ghost-in-the-Loop}, a humanoid teleoperation framework that enables switching between human-operated and AI-generated control while maintaining a consistent robotic embodiment. The system combines voice cloning, facial animation, and speech-driven gesture generation, allowing both control sources to be presented through the same visual and behavioral interface.

As an initial exploration, we conducted an online perception study ($N=50$) in which participants viewed short video clips generated using the system and judged whether the robot was controlled by a human or an AI. Our results suggest that observers often struggled to distinguish between the two control sources in brief interactions. Qualitative responses further indicate that participants relied primarily on perceived naturalness, temporal coordination, and consistency across speech, facial expression, and gesture when forming these judgments.

\section{Ghost-in-the-Loop}

\textit{Ghost-in-the-Loop} is a humanoid teleoperation framework that supports switching between human-operated and AI-generated behavior while maintaining a consistent robotic embodiment. The system is implemented on a Unitree G1 humanoid robot equipped with a head-mounted facial display and onboard camera. Human and AI control share the same output channels, for example, voice, facial expression, and upper-body motion, allowing both control sources to be presented through a visually consistent interface.

The framework consists of three components (\autoref{fig:ghost_arch}): \textsc{Local}, \textsc{Remote}, and \textsc{AI}. The \textsc{Local} side contains the humanoid robot and participant-facing interface. The \textsc{Remote} side enables a human operator to observe the interaction and teleoperate the robot. The \textsc{AI} side generates speech, facial expressions, and gestures using a collection of generative models.

For human-operated behavior, speech is transmitted through LiveKit~\cite{livekit}, facial expressions are captured using MediaPipe FaceLandmarker~\cite{mediapipe_face_landmarker} and converted into an Apple ARKit-compatible blendshape representation~\cite{apple_arkit}, and upper-body motion is teleoperated through a Meta Quest 3-based setup~\cite{unitree_xr_teleoperate}.

\begin{figure}[h]
\centering
\includegraphics[width=\linewidth]{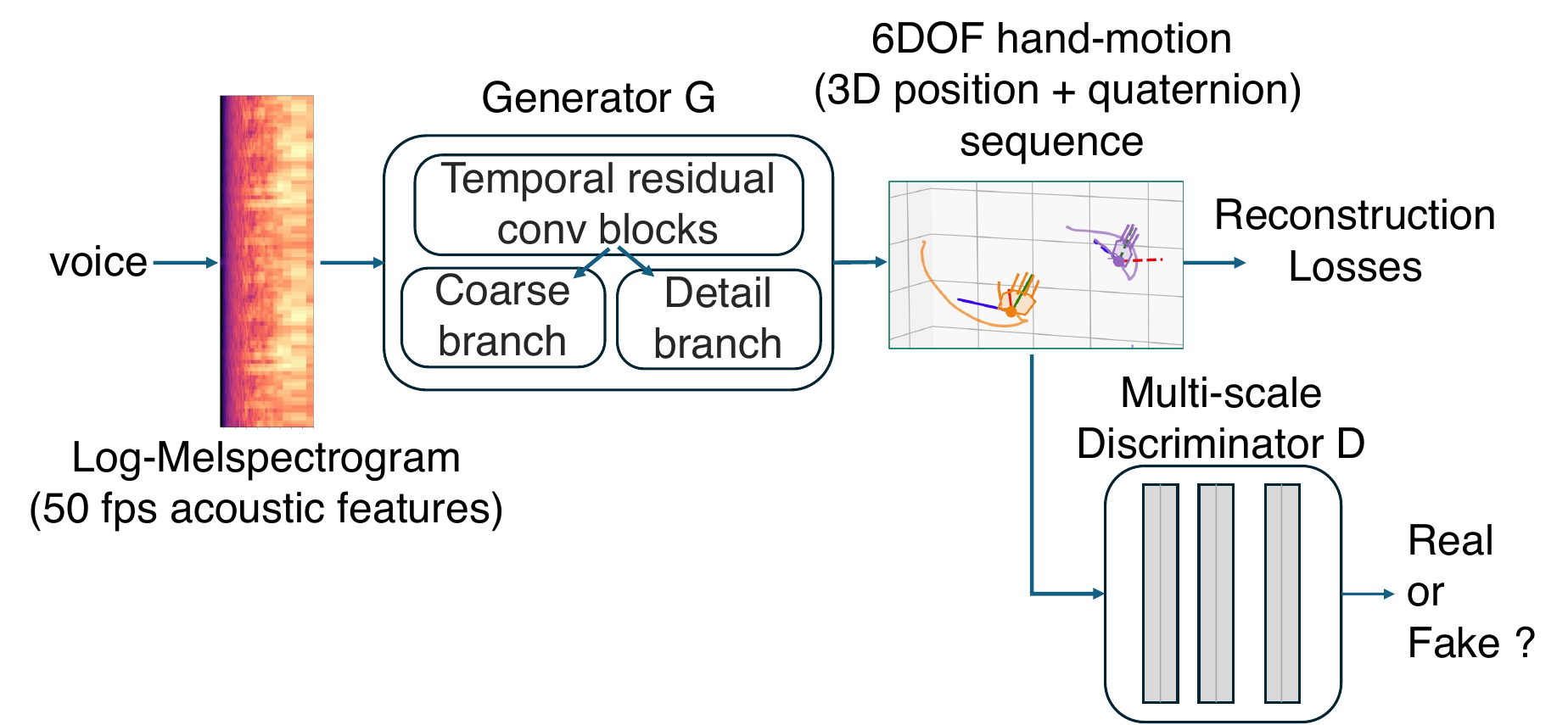}
\caption{\textit{voice2hand}: generating hand gestures from speech.}
\Description{A diagram of a speech-to-gesture model. Audio input is converted into a log-Mel spectrogram and fed into a generator network composed of temporal convolutional blocks with coarse and detail branches. The generator outputs left- and right-hand 6DoF motion sequences. A multi-scale discriminator evaluates the generated motion as real or fake, and a reconstruction loss is applied during training.}
\label{fig:voice2hand_model}
\end{figure}

For AI-generated behavior, participant speech is processed using the OpenAI Realtime API~\cite{openai_realtime_api}. Responses are synthesized with a cloned target voice using ElevenLabs~\cite{elevenlabs_voice_cloning}, facial animations are generated using NVIDIA Audio2Face~\cite{nvidia_audio2face_nim}, and gestures are generated using voice2hand (\autoref{fig:voice2hand_model}), a speech-driven hand-motion model derived from Speech2Gesture~\cite{ginosar2019gestures}. All outputs are rendered through the same presentation pipeline used in human teleoperation.

\section{Preliminary Evaluation}

To explore whether people can distinguish between human and AI agency in humanoid teleoperation, we conducted a preliminary online perception study using interaction clips. We recruited 50 participants through Prolific. To generate the study stimuli, we first used \textit{Ghost-in-the-Loop} to record 8 short humanoid interaction clips under 2 control conditions. In the \textbf{Human Operator} condition, the robot was directly controlled by a human teleoperator. In the \textbf{AI Control} condition, the robot was controlled by the AI pipeline, which generated speech, facial expressions, and gestures while maintaining the same embodied appearance and presentation channels. Across both conditions, the same humanoid robot, facial identity, voice identity, and recording environment were used.

The 8 clips covered both task-oriented and casual conversational scenarios, with durations ranging from approximately 5 to 40 seconds. Participants viewed each clip individually and rated the perceived source of control on a continuous scale ranging from $-10$ (AI) to $10$ (Human). Participants were also asked to provide a brief explanation describing how they arrived at their judgment~\cite{li2026we}.


\subsection{Findings}

Participants showed limited ability to distinguish between the Human Operator and AI Control conditions. The mean rating difference between conditions was small ($0.34$ on a $[-10,10]$ scale), with a 90\% confidence interval of $[-0.98,1.67]$. A TOST analysis using equivalence bounds of $\pm2$ points suggested practical equivalence ($p=.021$), indicating that the two control conditions were difficult to distinguish in this short-video setting.

Qualitative responses revealed three recurring themes. First, participants frequently relied on the overall naturalness of speech and behavior when making judgments. Second, temporal cues such as pacing, response timing, and movement regularity strongly influenced perceived agency. Third, participants often referred to the consistency between speech, facial expression, and gesture. Clips were more likely to be judged as AI-controlled when participants perceived rigid timing, repetitive movements, or mismatches across modalities, whereas clips exhibiting coherent multimodal behavior were often judged as human-operated.

Overall, these findings suggest that human and AI agency can be difficult to distinguish when presented through the same embodied interface, particularly in brief and structured interaction scenarios.


\section{Conclusion and Future Work}

We presented \textit{Ghost-in-the-Loop}, a humanoid teleoperation framework that enables both human-operated and AI-generated behavior to be presented through the same robotic embodiment. A preliminary study ($N=50$) suggests that people often struggle to distinguish between human and AI agency in short interactions. Looking forward, we are interested in understanding how agency is perceived during longer and more interactive encounters, and how future telepresence systems might dynamically combine human and AI control while preserving identity, transparency, and trust.

\begin{acks}
This work was supported by JST Moonshot R\&D Grant JPMJMS2012 and JST ASPIRE JPMJAP2503.
\end{acks}

\bibliographystyle{ACM-Reference-Format}
\bibliography{reference}





\end{document}